\documentclass{llncs}
\usepackage{amsfonts}
\input tcilatex

\begin{document}

\author{Zining Cao \\
\institute{
           Department of Computer Science and Technology \\
Nanjing University of Aero. \& Astro. \\ Nanjing 210016, P. R. China \\
    \email{caozn@nuaa.edu.cn}}}
\title{Reducing Higher Order $\pi $-Calculus to Spatial Logics}
\maketitle

\begin{abstract}
In this paper, we show that the theory of processes can be reduced to the
theory of spatial logic. Firstly, we propose a spatial logic $SL$ for higher
order $\pi $-calculus, and give an inference system of $SL$. The soundness
and incompleteness of $SL$ are proved. Furthermore, we show that the
structure congruence relation and one-step transition relation can be
described as the logical relation of $SL$ formulae. We also extend
bisimulations for processes to that for $SL$ formulae. Then we extend all
definitions and results of $SL$ to a weak semantics version of $SL,$ called $%
WL.$ At last, we add $\mu $-operator to $SL.$ This new logic is named $\mu
SL.$ We show that $WL$ is a sublogic of $\mu SL$ and replication operator
can be expressed in $\mu SL.$
\end{abstract}

\section{Introduction}

Higher order $\pi $-calculus was proposed and studied intensively in
Sangiorgi's dissertation \cite{San92}. In higher order $\pi $-calculus,
processes and abstractions over processes of arbitrarily high order, can be
communicated. Some interesting equivalences for higher order $\pi $%
-calculus, such as barbed equivalence, context bisimulation and normal
bisimulation, were presented in \cite{San92}. Barbed equivalence can be
regarded as a uniform definition of bisimulation for a variety of concurrent
calculi. Context bisimulation is a very intuitive definition of bisimulation
for higher order $\pi $-calculus, but it is heavy to handle, due to the
appearance of universal quantifications in its definition. In the definition
of normal bisimulation, all universal quantifications disappeared, therefore
normal bisimulation is a very economic characterization of bisimulation for
higher order $\pi $-calculus. The coincidence between the three weak
equivalences was proven \cite{San92,San96,JR03}. Moreover, this proposition
was generalized to the strong case \cite{Cao06}.

Spatial logic was presented in \cite{CC03}. Spatial logic extends classical
logic with connectives to reason about the structure of the processes. The
additional connectives belong to two families. Intensional operators allow
one to inspect the structure of the process. A formula $A_1|A_2$ is
satisfied whenever we can split the process into two parts satisfying the
corresponding subformula $A_i$, $i=1,2$. In the presence of restriction in
the underlying model, a process $P$ satisfies formula $n\circledR A$ if we
can write $P$ as $(\nu n)P^{\prime }$ with $P^{\prime }$ satisfying $A$.
Finally, formula $0$ is only satisfied by the inaction process. Connectives $%
|$ and $\circledR$ come with adjunct operators, called guarantee ($%
\triangleright $) and hiding ($\oslash $) respectively, that allow one to
extend the process being observed. In this sense, these can be called
contextual operators. $P$ satisfies $A_1\triangleright A_2$ whenever the
spatial composition (using $|$) of $P$ with any process satisfying $A_1$
satisfies $A_2$, and $P$ satisfies $A\oslash n$ if $(\nu n)P$ satisfies $A$.
Some spatial logics have an operator for fresh name quantification \cite
{CC04}.

There are lots of works of spatial logics for $\pi $-calculus and {\it %
Mobile Ambients}. In some papers, spatial logic was studied on its relations
with structural congruence, bisimulation, model checking and type system of
process calculi \cite{Cai08,Cai07,Cao06a,CG00,San01}.

The main idea of this paper is that the theory of processes can be reduced
to the theory of spatial logic.

In this paper, we present a spatial logic for higher order $\pi $-calculus,
called $SL$, which comprises some action temporal operators such as $\langle
\tau \rangle \ $and $\langle a\langle A\rangle \rangle ,$ some spatial
operators such as prefix and composition, some adjunct operators of spatial
operators such as $\triangleright $ and $\oslash ,$ and some operators on
the property of free names and bound names such as $\ominus n$ and $%
\widetilde{\ominus }.$ We give an inference system of $SL$, and prove the
soundness of the inference system for $SL$. Furthermore, we show that there
is no finite complete inference system for $SL.$ Then we study the relation
between processes and $SL$ formulas. We show that a $SL$ formula can be
viewed as a specification of processes, and conversely, a process can be
viewed as a special kind of $SL$ formulas. Therefore, $SL$ is a
generalization of processes, which extend process with specification
statements. We show that the structural congruence relation and one-step
transition relation can be described as the logical relation of $SL$
formulas. We also show that bisimulations for higher order processes
coincides with logical equivalence with respect to some fragment of a
sublogic of $SL$.

Furthermore, we give a weak semantics version of $SL,$ called $WL$, where
the internal action is unobservable. The results of $SL$ are extended to $WL$%
, such as an inference system for $WL$, the soundness of this inference
system, and no finite complete inference system for $WL$.\

Finally, we add $\mu $-operator to $SL.$ The new logic is named $\mu SL.$ We
show that $WL$ is a sublogic of $\mu SL$ and replication operator can be
expressed in $\mu SL.$ Thus $\mu SL$ is a powerful logic which can express
both strong semantics and weak semantics for higher order processes.

This paper is organized as follows: In Section 2, we briefly review higher
order $\pi $-calculus. In Section 3, we present a spatial logic $SL$,
including its syntax, semantics and inference system. The soundness and
incompleteness of the inference system of $SL$\ are proved. Furthermore, we
discuss that $SL$ can be regarded as a specification language of processes
and processes can be regarded as a kind of special formulas of $SL$.
Bisimulation in higher order $\pi $-calculus coincides with logical
equivalence with respect to some fragment of a sublogic of $SL$. In Section
4, we give a weak semantics version of $SL$, called $WL$. We generalize
concepts and results of $SL$\ to $WL$. In Section 5, we add $\mu $-operator
to $SL.$ The new logic is named $\mu SL.$ We studied the expressive power of
this extension. The paper is concluded in Section 6.

\section{Higher Order $\pi $-Calculus}

\subsection{Syntax and Labelled Transition System}

In this section we briefly recall the syntax and labelled transition system
of the higher order $\pi $-calculus. Similar to \cite{San96}, we only focus
on a second-order fragment of the higher order $\pi $-calculus, i.e., there
is no abstraction in this fragment.

We assume a set $N$ of names, ranged over by $a,b,c,...$ and a set $Var$ of
process variables, ranged over by $X,Y,Z,U,...$. We use $E,F,P,Q,...$ to
stand for processes. $Pr$ denotes the set of all processes.

We first give the grammar for the higher order $\pi $-calculus processes as
follows:

$P::=0\ |\ U\ |\ \pi .P\ |\ P_1|P_2\ |\ (\nu a)P$

$\pi $ is called a prefix and can have one of the following forms:

$\pi ::=$ $a(U)$ $|$ $\overline{a}\langle P\rangle ,$ here $a(U)$ is a
higher order input prefix and $\overline{a}\langle P\rangle $ is a higher
order output prefix.

In each process of the form $(\nu a)P$ the occurrence of $a$ is bound within
the scope of $P$. An occurrence of $a$ in a process is said to be free iff
it does not lie within the scope of a bound occurrence of $a$. The set of
names occurring free in $P$ is denoted $fn(P)$. An occurrence of a name in a
process is said to be bound if it is not free, we write the set of bound
names as $bn(P)$. $n(P)$ denotes the set of names of $P$, i.e., $%
n(P)=fn(P)\cup bn(P)$. The definition of substitution in process terms may
involve renaming of bound names when necessary to avoid name capture.

Higher order input prefix $a(U).P$ binds all free occurrences of $U$ in $P$.
The set of variables occurring free in $P$ is denoted $fv(P)$. We write the
set of bound variables as $bv(P)$. A process is closed if it has no free
variable; it is open if it may have free variables. $Pr^c$ is the set of all
closed processes.

Processes $P$ and $Q$ are $\alpha $-convertible, $P\equiv _\alpha Q$, if $Q$
can be obtained from $P$ by a finite number of changes of bound names and
variables. For example, $(\nu b)(\overline{a}\langle b(U).U\rangle .0)\equiv
_\alpha (\nu c)(\overline{a}\langle c(U).U\rangle .0)$.

Structural congruence is the smallest congruence relation that validates the
following axioms: $P|Q\equiv Q|P;$ $(P|Q)|R\equiv P|(Q|R);$ $P|0\equiv P;$ $%
(\nu a)0\equiv 0;$ $(\nu m)(\nu n)P\equiv (\nu n)(\nu m)P;$ $(\nu
a)(P|Q)\equiv P|(\nu a)Q\ $if $a\notin fn(P).$

In \cite{Par01}, Parrow has shown that in higher order $\pi $-calculus, the
replication can be defined by other operators such as higher order prefix,
parallel and restriction. For example, $!P$ can be simulated by $R_P=(\nu
a)(D|\overline{a}\langle P|D\rangle .0)$, where $D=a(X).(X|\overline{a}%
\langle X\rangle .0).$

The operational semantics of higher order processes is given in Table 1. We
have omitted the symmetric cases of the parallelism and communication rules.

\begin{center}
$ALP:\dfrac{P\stackrel{\alpha }{\longrightarrow }P^{\prime }}{Q\stackrel{%
\alpha }{\longrightarrow }Q^{\prime }}$ $P\equiv Q,P^{\prime }\equiv
Q^{\prime }$

$OUT:\overline{a}\langle E\rangle .P\stackrel{\overline{a}\langle E\rangle }{%
\longrightarrow }P$

$IN:a(U).P\stackrel{a\langle E\rangle }{\longrightarrow }P\{E/U\}$ $%
bn(E)=\emptyset $

$PAR:\dfrac{P\stackrel{\alpha }{\longrightarrow }P^{\prime }}{P|Q\stackrel{%
\alpha }{\longrightarrow }P^{\prime }|Q}$ $bn(\alpha )\cap fn(Q)=\emptyset $

$COM:\dfrac{P\stackrel{(\nu \widetilde{b})\overline{a}\langle E\rangle }{%
\longrightarrow }P^{\prime }\quad Q\stackrel{a\langle E\rangle }{%
\longrightarrow }Q^{\prime }}{P|Q\stackrel{\tau }{\longrightarrow }(\nu
\widetilde{b})(P^{\prime }|Q^{\prime })}$ $\widetilde{b}\cap fn(Q)=\emptyset
$

$RES:\dfrac{P\stackrel{\alpha }{\longrightarrow }P^{\prime }}{(\nu a)P%
\stackrel{\alpha }{\longrightarrow }(\nu a)P^{\prime }}$ $a\notin n(\alpha )$

$OPEN:\dfrac{P\stackrel{(\nu \widetilde{c})\overline{a}\langle E\rangle }{%
\longrightarrow }P^{\prime }}{(\nu b)P\stackrel{(\nu b,\widetilde{c})%
\overline{a}\langle E\rangle }{\longrightarrow }P^{\prime }}$ $a\neq b,\
b\in fn(E)-\widetilde{c}$

Table 1. The operational semantics of higher order $\pi $-calculus
\end{center}

\subsection{Bisimulations in Higher Order $\pi $-Calculus}

Context bisimulation and contextual barbed bisimulation were presented in
\cite{San92,San96} to describe the behavioral equivalences for higher order $%
\pi $-calculus. Let us review the definition of these bisimulations. In the
following, we abbreviate $P\{E/U\}$ as $P\langle E\rangle $.

Context bisimulation is an intuitive definition of bisimulation for higher
order $\pi $-calculus.

{\bf Definition 1} A symmetric relation $R\subseteq Pr^c\times Pr^c$ is a
strong context bisimulation if $P\ R\ Q$ implies:

(1) whenever $P\stackrel{\tau }{\longrightarrow }P^{\prime }$, there exists $%
Q^{\prime }$ such that $Q\stackrel{\tau }{\longrightarrow }Q^{\prime }$ and $%
P^{\prime }$ $R$ $Q^{\prime }$;

(3) whenever $P\stackrel{a\langle E\rangle }{\longrightarrow }P^{\prime }$,
there exists $Q^{\prime }$ such that $Q\stackrel{a\langle E\rangle }{%
\longrightarrow }Q^{\prime }$ and $P^{\prime }$ $R$ $Q^{\prime }$;

(4) whenever $P\stackrel{(\nu \widetilde{b})\overline{a}\langle E\rangle }{%
\longrightarrow }P^{\prime }$, there exist $Q^{\prime }$, $F$, $\widetilde{c}
$ such that $Q\stackrel{(\nu \widetilde{c})\overline{a}\langle F\rangle }{%
\longrightarrow }Q^{\prime }$ and for all $C(U)$ with $fn(C(U))\cap \{%
\widetilde{b},\widetilde{c}\}=\emptyset $, $(\nu \widetilde{b})(P^{\prime
}|C\langle E\rangle )$ $R$ $(\nu \widetilde{c})(Q^{\prime }|C\langle
F\rangle )$. Here $C(U)$ represents a process containing a unique free
variable $U.$

We write $P\sim _{Ct}Q$ if $P$ and $Q$ are strongly context bisimilar.

Contextual barbed equivalence can be regarded as a uniform definition of
bisimulation for a variety of process calculi.

{\bf Definition 2} A symmetric relation $R\subseteq Pr^c\times Pr^c$ is a
strong contextual barbed bisimulation if $P$ $R$ $Q$ implies:

(1) $P|C$ $R$ $Q|C$ for any $C;$

(2) whenever $P\stackrel{\tau }{\longrightarrow }P^{\prime }$ then there
exists $Q^{\prime }$ such that $Q\stackrel{\tau }{\longrightarrow }Q^{\prime
}$ and $P^{\prime }$ $R$ $Q^{\prime }$;

(3) $P\downarrow _\mu $ implies $Q\downarrow _\mu $, where $P\downarrow _a$
if $\exists P^{\prime },$ $P\stackrel{a\langle E\rangle }{\longrightarrow }%
P^{\prime },$ and $P\downarrow _{\overline{a}}$ if $\exists P^{\prime },$ $P%
\stackrel{(\nu \widetilde{b})\overline{a}\langle E\rangle }{\longrightarrow }%
P^{\prime }.$

We write $P\sim _{Ba}Q$ if $P$ and $Q$ are strongly contextual barbed
bisimilar.

Intuitively, a tau action represents the internal action of processes. If we
just consider external actions, then we should adopt weak bisimulations to
characterize the equivalence of processes.

{\bf Definition 3} A symmetric relation $R\subseteq Pr^c\times Pr^c$ is a
weak context bisimulation if $P\ R\ Q$ implies:

(1) whenever $P\stackrel{\varepsilon }{\Longrightarrow }P^{\prime }$, there
exists $Q^{\prime }$ such that $Q\stackrel{\varepsilon }{\Longrightarrow }%
Q^{\prime }$ and $P^{\prime }$ $R$ $Q^{\prime }$;

(2) whenever $P\stackrel{a\langle E\rangle }{\Longrightarrow }P^{\prime }$,
there exists $Q^{\prime }$ such that $Q\stackrel{a\langle E\rangle }{%
\Longrightarrow }Q^{\prime }$ and $P^{\prime }$ $R$ $Q^{\prime }$;

(3) whenever $P\stackrel{(\nu \widetilde{b})\overline{a}\langle E\rangle }{%
\Longrightarrow }P^{\prime }$, there exist $Q^{\prime }$, $F$, $\widetilde{c}
$ such that $Q\stackrel{(\nu \widetilde{c})\overline{a}\langle F\rangle }{%
\Longrightarrow }Q^{\prime }$ and for all $C(U)$ with $fn(C(U))\cap \{%
\widetilde{b},\widetilde{c}\}=\emptyset $, $(\nu \widetilde{b})(P^{\prime
}|C\langle E\rangle )$ $R$ $(\nu \widetilde{c})(Q^{\prime }|C\langle
F\rangle )$. Here $C(U)$ represents a process containing a unique free
variable $U.$

We write $P\approx _{Ct}Q$ if $P$ and $Q$ are weakly context bisimilar.

{\bf Definition 4} A symmetric relation $R\subseteq Pr^c\times Pr^c$ is a
weak contextual barbed bisimulation if $P$ $R$ $Q$ implies:

(1) $P|C$ $R$ $Q|C$ for any $C;$

(2) whenever $P\stackrel{\varepsilon }{\Longrightarrow }P^{\prime }$ then
there exists $Q^{\prime }$ such that $Q\stackrel{\varepsilon }{%
\Longrightarrow }Q^{\prime }$ and $P^{\prime }$ $R$ $Q^{\prime }$;

(3) $P\Downarrow _\mu $ implies $Q\Downarrow _\mu $, where $P\Downarrow _\mu
$ if $\exists P^{\prime },$ $P\stackrel{\varepsilon }{\Longrightarrow }%
P^{\prime }$ and $P^{\prime }\downarrow _\mu .$

We write $P\approx _{Ba}Q$ if $P$ and $Q$ are weakly contextual barbed
bisimilar.

\section{Logics for Strong Semantics}

In this section, we present a logic to reason about higher order $\pi $%
-calculus called $SL$. This logic extends propositional logic with three
kinds of connectives: action temporal operators, spatial operators,
operators about names and variables. We give the syntax and semantics of $%
SL. $ The inference system of $SL$ is also given. We prove the soundness and
incompleteness of this inference system. As far as we know, this is the
first result on the completeness problem of the inference system of spatial
logic. Furthermore, we show that structural congruence, one-step transition
relation and bisimulation can all be characterized by this spatial logic. It
is well known that structural congruence, one-step transition relation and
bisimulation are the central concepts in the theory of processes, and almost
all the studies of process calculi are about these concepts. Therefore, our
study gives an approach of reducing theory of processes to theory of spatial
logic. Moreover, since processes can be regarded as a special kind of
spatial logic formulas, spatial logic can be viewed as an extension of
process calculus. Based on spatial logic, it is possible to propose a
refinement calculus \cite{MGRV94} of concurrent processes.

\subsection{Syntax and Semantics of Logic $SL$}

Now we introduce a logic called $SL,$ which is a spatial logic for higher
order $\pi $-calculus.

{\bf Definition 5} Syntax of logic $SL$

$A::=\top |$ $\bot |$ $\neg A$ \TEXTsymbol{\vert} $A_1\wedge A_2$
\TEXTsymbol{\vert} $\langle \tau \rangle A$ \TEXTsymbol{\vert} $\langle
a\langle A_1\rangle \rangle A_2$ \TEXTsymbol{\vert} $\langle a[A_1]\rangle
A_2$ \TEXTsymbol{\vert} $\langle \overline{a}\langle A_1\rangle \rangle A_2$
\TEXTsymbol{\vert} $0$ $|$ $X$ $|$ $a\odot X.A$ \TEXTsymbol{\vert} $%
A\setminus a\odot X$ \TEXTsymbol{\vert} $\overline{a}\langle A_1\rangle .A_2$
\TEXTsymbol{\vert} $A\setminus \overline{a}$ \TEXTsymbol{\vert} $A_1|A_2$
\TEXTsymbol{\vert} $A_1\triangleright A_2$ \TEXTsymbol{\vert} $a\circledR A$
\TEXTsymbol{\vert} $A\oslash a$ \TEXTsymbol{\vert} $({\bf N}x)A$ $|$ $({\bf N%
}X)A$ $|$ $(\ominus a)A$ \TEXTsymbol{\vert} $(\tilde \ominus )A$ \TEXTsymbol{%
\vert} $a\not =b$

In $({\bf N}x)A,$ $({\bf N}X)A,$ the variables $x$ (and $X$) are bound with
scope the formula $A$. We assume defined on formulas the standard relation $%
\equiv _\alpha $ of $\alpha $-conversion (safe renaming of bound variables),
but we never implicitly take formulas ``up to $\alpha $-conversion'': our
manipulation of variables via $\alpha $-conversion steps is always quite
explicit. The set $fn(A)$ of free names in $A,$ and the set $fpv(A)$ of free
propositional variables in $A,$ are defined in the usual way. A formula is
closed if it has no free variable such as $X$, it is open if it may have
free variables. $SL^c$ is the set of all closed formulas. In the following,
we use $A\{b/a\}$ to denote the formula obtained by replacing all occurrence
of $a$ in $A$ by $b.$ Similarly, we use $A\{Y/X\}$ to denote the formula
obtained by replacing all occurrence of $Y$ in $A$ by $X.$ It is easy to see
that a process can also be regarded as a spatial formula. For example,
process $\overline{a}\langle E\rangle .P$ is also a spatial formula. In this
paper, we say that such a formula is in the form of process formula.

{\bf Definition 6} Semantics of logic $SL$

$[[\top ]]_{Pr}=Pr$

$[[\bot ]]_{Pr}=\emptyset $

$[[\neg A]]_{Pr}=Pr-[[A]]_{Pr}$

$[[A_1\wedge A_2]]_{Pr}=[[A_1]]_{Pr}\cap [[A_2]]_{Pr}$

$[[\langle \tau \rangle A]]_{Pr}=\{P$ $|$ $\exists Q$. $P\stackrel{\tau }{%
\longrightarrow }Q$ and $Q\in [[A]]_{Pr}\}$

$[[\langle a\langle A_1\rangle \rangle A_2]]_{Pr}=\{P$ $|$ $\exists P_1,P_2$%
. $P\stackrel{a\langle P_1\rangle }{\longrightarrow }P_2,$ $P_1\in
[[A_1]]_{Pr}$ and $P_2\in [[A_2]]_{Pr}\}$

$[[\langle a[A_1]\rangle A_2]]_{Pr}=\{P$ $|$ $\forall R,R\in
[[A_1]]_{Pr},\exists Q$. $P\stackrel{a\langle R\rangle }{\longrightarrow }Q$
and $Q\in [[A_2]]_{Pr}\}$

$[[\langle \overline{a}\langle A_1\rangle \rangle A_2]]_{Pr}=\{P$ $|$ $%
\exists P_1,P_2$. $P\stackrel{(\nu \widetilde{b})\overline{a}\langle
P_1\rangle }{\longrightarrow }P_2,$ $(\nu \widetilde{b})P_1\in [[A_1]]_{Pr}$
and $P_2\in [[A_2]]_{Pr}\}$

$[[0]]_{Pr}=\{P$ $|$ $P\equiv 0\}$

$[[X]]_{Pr}=\{P$ $|$ $P\equiv X\}$

$[[a\odot X.A]]_{Pr}=\{P$ $|$ $\exists Q$. $P\equiv a(X).Q$ and $Q\in
[[A]]_{Pr}\}$

$[[A\setminus a\odot X]]_{Pr}=\{P$ $|$ $a(X).P\in [[A]]_{Pr}\}$

$[[\overline{a}\langle A_1\rangle .A_2]]_{Pr}=\{P$ $|$ $\exists P_1,P_2$. $%
P\equiv \overline{a}\langle P_1\rangle .P_2$, $P_1\in [[A_1]]_{Pr}$ and $%
P_2\in [[A_2]]_{Pr}\}$

$[[A\setminus \overline{a}]]_{Pr}=\{P$ $|$ $\overline{a}\langle P\rangle
.0\in [[A]]_{Pr}\}$

$[[A_1|A_2]]_{Pr}=\{P$ $|$ $\exists Q_1,Q_2$. $P\equiv Q_1|Q_2$, $Q_1\in
[[A_1]]_{Pr}$ and $Q_2\in [[A_2]]_{Pr}\}$

$[[A_1\triangleright A_2]]_{Pr}=\{P$ $|$ $\forall Q$. $Q\in [[A_1]]_{Pr}$
implies $P|Q\in [[A_2]]_{Pr}\}$

$[[a\circledR A]]_{Pr}=\{P$ $|$ $\exists Q$. $P\equiv (\nu a)Q$ and $Q\in
[[A]]_{Pr}\}$

$[[A\oslash a]]_{Pr}=\{P$ $|$ $(\nu a)P\in [[A]]_{Pr}\}$

$[[({\bf N}x)A]]_{Pr}=\cup _{n\notin fn(({\bf N}x)A)}([[A\{n/x\}]]_{Pr}%
\backslash \{P$ $|$ $n\in fn(P)\})$

$[[({\bf N}X)A]]_{Pr}=\cup _{V\notin fpv(({\bf N}X)A)}([[A\{V/X\}]]_{Pr}%
\backslash \{P$ $|$ $V\in fpv(P)\})$

$[[(\ominus a)A]]_{Pr}=\{P$ $|$ $a\notin fn(P)$ and $P\in [[A]]_{Pr}\}$

$[[(\tilde \ominus )A]]_{Pr}=\{P$ $|$ $\exists Q$. $P\equiv Q$ and $%
bn(Q)=\emptyset $ and $Q\in [[A]]_{Pr}\}$

$[[a\neq b]]_{Pr}=Pr$ if $a\neq b$

$[[a\neq b]]_{Pr}=\emptyset $ if $a=b$

In $SL$, formula $\langle a\langle A_1\rangle \rangle A_2$ is satisfied by
the processes that can receive a process satisfying $A_1$ and then become a
process satisfying $A_2.$ Formula $\langle a[A_1]\rangle A_2$ is satisfied
by processes that if it receive any process satisfying $A_1$ then it becomes
a process satisfying $A_2.$ $A\setminus a\odot X$ is an adjunct operator of $%
a\odot X.A,$ and $A\setminus \overline{a}$ is an adjunct operator of $%
\overline{a}\langle A\rangle .0.$ $(\ominus a)A$ is satisfied by processes
that satisfies $A$ and $a$ is not its free name. $(\tilde \ominus )A$ is
satisfied by processes that satisfy $A$ and have no bound names. Other
operators in $SL$ are well known in spatial logic or can be interpreted
similarly as above operators.

{\bf Definition 7} $P\models _{SL}A$ if $P\in [[A]]_{Pr}.$

{\bf Definition 8} For a set of formulas $\Gamma $ and a formula $A$, we
write $\Gamma \models _{SL}A$, if $A$ is valid in all processes that satisfy
all formulas of $\Gamma $.

{\bf Definition 9} If ``$A_1,...,A_n$ infer $B$'' is an instance of an
inference rule, and if the formulas $A_1,...,A_n$ have appeared earlier in
the proof, then we say that $B$ follows from an application of an inference
rule. A proof is said to be from $\Gamma $ to $A$ if the premise is $\Gamma $
and the last formula is $A$ in the proof. We say $A$ is provable from $%
\Gamma $ in an inference system $AX$, and write $\Gamma \vdash _{AX}A$, if
there is a proof from $\Gamma $ to $A$ in $AX$.

For example, the following sets can be defined by operators in $SL$:

$\{P$ $|$ $\forall P_1$. $P_1\in [[A_1]]_{Pr}$ implies $\overline{a}\langle
P_1\rangle .P\in [[A_2]]_{Pr}\}=[[(b\odot Y.\overline{a}\langle A_1\rangle
.Y\triangleright \langle \tau \rangle A_2)\setminus \overline{b}]]_{Pr}$

$\{P$ $|$ $\forall P_1$. $P_1\in [[A_1]]_{Pr}$ implies $\overline{a}\langle
P\rangle .P_1\in [[A_2]]_{Pr}\}=[[(b\odot Y.\overline{a}\langle Y\rangle
.A_1\triangleright \langle \tau \rangle A_2)\setminus \overline{b}]]_{Pr}$

$\{P$ $|$ $a\in fn(P)$ and $P\in [[A]]_{Pr}\}=[[\neg (\ominus a)\top \wedge
A]]_{Pr}$

$\{P$ $|$ $X\in fv(P)$ and $P\in [[A]]_{Pr}\}=[[\neg (\ominus X)\top \wedge
A]]_{Pr}$

$({\bf H}x)A=({\bf N}x)x\circledR A,$ which is related to name restriction
in an appropriate way; namely, that if process $P$ satisfies formulas $%
A\{n/x\},$ then $(\nu n)P$ satisfies $({\bf H}x)A.$

$(a{\bf H}X)A=({\bf N}X)a\odot X.A,$ which is related to process variable
restriction in an appropriate way; namely, that if process $P$ satisfies
formulas $A\{U/X\},$ then $a(U).P$ satisfies $(a{\bf H}X)A.$

\subsection{Inference System of $SL$}

Now we list a number of valid properties of spatial logic. The combination
of the complete inference system of first order logic and the following
axioms and rules form the inference system $S$ of $SL$.

$$
\begin{array}{l}
\langle \alpha \rangle \bot \rightarrow \bot \\
a\odot X.\bot \rightarrow \bot \\
\overline{a}\langle \top \rangle .\bot \rightarrow \bot \\ \overline{a}%
\langle \bot \rangle .\top \rightarrow \bot \\ \bot \setminus a\odot
X\rightarrow \bot \\
\bot \setminus
\overline{a}\rightarrow \bot \\ A|\bot \rightarrow \bot \\
A\triangleright \bot \rightarrow \neg A
\end{array}
\quad \quad \quad
\begin{array}{l}
\bot \triangleright A\leftrightarrow \top \\
a
\circledR \bot \rightarrow \bot \\ \bot \oslash a\rightarrow \bot \\
(\ominus a)\bot \rightarrow \bot \\
(
{\bf N}x)\bot \rightarrow \bot \\ (\tilde \ominus )\bot \rightarrow \bot \\
(
{\bf N}X)\bot \rightarrow \bot \\ A|B\leftrightarrow B|A
\end{array}
\quad \quad \quad
\begin{array}{l}
(A|B)|C\leftrightarrow A|(B|C) \\
A|0\leftrightarrow A \\
a
\circledR 0\leftrightarrow 0 \\ a
\circledR b\circledR A\leftrightarrow b\circledR a\circledR A \\ a
\circledR ((\ominus a)A|B)\leftrightarrow (\ominus a)A|a\circledR B \\ a
\circledR A\rightarrow ({\bf N}b)b\circledR A\{b/a\} \\ a\odot
X.A\rightarrow (
{\bf N}Y)a\odot Y.A\{Y/X\} \\ (\ominus a)0\leftrightarrow 0
\end{array}
$$

$$
\begin{array}{l}
(\ominus a)X\leftrightarrow X \\
(\ominus a)a\odot X.A\leftrightarrow \bot  \\
(\ominus a)
\overline{a}\langle B\rangle .A\leftrightarrow \bot  \\ a\neq b\rightarrow
((\ominus a)b\odot X.A\leftrightarrow b\odot X.(\ominus a)A) \\
a\neq b\rightarrow ((\ominus a)
\overline{b}\langle B\rangle .A\leftrightarrow \overline{b}\langle (\ominus
a)B\rangle .(\ominus a)A) \\ (\ominus a)A|(\ominus a)B\leftrightarrow
(\ominus a)(A|B) \\
a\neq b\rightarrow ((\ominus a)(\ominus b)A\leftrightarrow (\ominus
b)(\ominus a)A) \\
(\ominus a)a\circledR A\leftrightarrow a\circledR A
\end{array}
\quad
\begin{array}{l}
(\tilde \ominus )0\leftrightarrow 0 \\
(\tilde \ominus )X\leftrightarrow X \\
(\tilde \ominus )a\odot X.A\leftrightarrow a\odot X.(\tilde \ominus )A \\
(\tilde \ominus )
\overline{a}\langle B\rangle .A\leftrightarrow \overline{a}\langle (\tilde
\ominus )B\rangle .(\tilde \ominus )A \\ (\tilde \ominus )A|(\tilde \ominus
)B\leftrightarrow (\tilde \ominus )(A|B) \\
(\tilde \ominus )a
\circledR \neg (\ominus a)\top \rightarrow \bot  \\ (
{\bf N}x)0\leftrightarrow 0 \\ ({\bf N}x)X\leftrightarrow X
\end{array}
$$

\begin{center}
$
\begin{array}{l}
(
{\bf N}x)a\odot X.A\leftrightarrow a\odot X.({\bf N}x)(x\neq a\wedge A) \\ (
{\bf N}x)\overline{a}\langle B\rangle .A\rightarrow \overline{a}\langle (%
{\bf N}x)(x\neq a\wedge B)\rangle .({\bf N}x)(x\neq a\wedge A) \\ (
{\bf N}x)(A|B)\rightarrow ({\bf N}x)A|({\bf N}x)B \\ (
{\bf N}x)x\neq a\wedge a\circledR A\rightarrow a\circledR ({\bf N}x)A \\ (
{\bf N}X)0\leftrightarrow 0 \\ ({\bf N}X)X\rightarrow Y
\end{array}
\quad
\begin{array}{l}
(
{\bf N}X)a\odot Y.A\leftrightarrow a\odot Y.({\bf N}X)A \\ (
{\bf N}X)\overline{a}\langle B\rangle .A\rightarrow \overline{a}\langle (%
{\bf N}X)B\rangle .({\bf N}X)A \\ (
{\bf N}X)(A|B)\rightarrow ({\bf N}X)A|({\bf N}X)B \\ (
{\bf N}X)a\circledR A\leftrightarrow a\circledR ({\bf N}X)A \\ a\odot
X.(A\setminus a\odot X)\rightarrow A \\
A\rightarrow (a\odot X.A)\setminus a\odot X)
\end{array}
$
\end{center}

$$
\begin{array}{l}
\overline{a}\langle A\setminus \overline{a}\rangle .0\rightarrow A \\
A\rightarrow ((
\overline{a}\langle A\rangle .0)\setminus \overline{a}) \\
(A|A\triangleright B)\rightarrow B \\
A\rightarrow (B\triangleright A|B) \\
a
\circledR (A\oslash a)\rightarrow A \\ A\rightarrow (a
\circledR A\oslash a) \\ \langle \alpha \rangle A,A\rightarrow B\vdash
\langle \alpha \rangle B
\end{array}
\quad
\begin{array}{l}
a\odot X.A,A\rightarrow B\vdash a\odot X.B \\
\overline{a}\langle C\rangle .A,A\rightarrow B\vdash \overline{a}\langle
C\rangle .B \\ \overline{a}\langle B\rangle .A,B\rightarrow C\vdash
\overline{a}\langle C\rangle .A \\ \langle
\overline{a}\langle B\rangle \rangle A,C\rightarrow B\vdash \langle
\overline{a}\langle C\rangle \rangle A \\ \langle a[B]\rangle A,C\rightarrow
B\vdash \langle a[C]\rangle A \\
A\setminus a\odot X,A\rightarrow B\vdash B\setminus a\odot X \\
A\setminus \overline{a},A\rightarrow B\vdash B\setminus \overline{a}
\end{array}
\quad
\begin{array}{l}
A\rightarrow B\vdash A|C\rightarrow B|C \\
a
\circledR A,A\rightarrow B\vdash a\circledR B \\ (\ominus a)A,A\rightarrow
B\vdash (\ominus a)B \\
(\tilde \ominus )A,A\rightarrow B\vdash (\tilde \ominus )B \\
\overline{a}\langle B\rangle .A\rightarrow \langle \overline{a}\langle
B\rangle \rangle A \\ (\langle \tau \rangle A)|B\rightarrow \langle \tau
\rangle (A|B) \\
(\langle a\langle C\rangle \rangle A)|B\rightarrow \langle a\langle C\rangle
\rangle (A|B)
\end{array}
$$

$$
\begin{array}{l}
(a\odot U.A\wedge ((\tilde \ominus )B\leftrightarrow B))\rightarrow \langle
a[B]\rangle A\{B/U\} \\
(((\ominus b_1,...,\ominus b_n)B\leftrightarrow B)\wedge ((\tilde \ominus
)C\leftrightarrow C))\rightarrow \\
\quad \quad \quad \quad ((\langle
\overline{a}\langle b_1\circledR ...b_n\circledR C\rangle \rangle
A)|B\rightarrow \langle \overline{a}\langle b_1\circledR ...b_n\circledR %
C\rangle \rangle (A|B)) \\ (((\ominus b_1,...,\ominus b_n)B\leftrightarrow
B)\wedge ((\tilde \ominus )C\leftrightarrow C))\rightarrow \\
\quad \quad \quad \quad ((\langle
\overline{a}\langle b_1\circledR ...b_n\circledR C\rangle \rangle A)|\langle
a[C]\rangle B\rightarrow \langle \tau \rangle b_1\circledR ...b_n\circledR %
(A|B)) \\ (a\neq b\wedge ((\ominus a)B\leftrightarrow B)\wedge ((\tilde
\ominus )B\leftrightarrow B))\rightarrow (a
\circledR \langle b\langle B\rangle \rangle A\rightarrow \langle b\langle
B\rangle \rangle a\circledR A) \\ (\wedge _{i=1}^na\neq b_i\wedge a\neq
c\wedge ((\ominus a)B\leftrightarrow B)\wedge ((\tilde \ominus
)B\leftrightarrow B))\rightarrow \\
\quad \quad \quad \quad (a
\circledR \langle \overline{c}\langle b_1\circledR ...b_n\circledR B\rangle
\rangle A\rightarrow \langle \overline{c}\langle b_1\circledR ...b_n%
\circledR B\rangle \rangle a\circledR A) \\ (a\neq b\wedge \wedge
_{i=1}^nb\neq c_i\wedge \ (B\rightarrow \neg (\ominus b)\top )\wedge
((\tilde \ominus )B\leftrightarrow B))\rightarrow \\
\quad \quad \quad \quad (b
\circledR \langle \overline{a}\langle c_1\circledR ...c_n\circledR B\rangle
\rangle A\rightarrow \langle \overline{a}\langle b\circledR c_1\circledR %
...c_n\circledR B\rangle \rangle A) \\ \langle a[B]\rangle A\rightarrow
\langle a\langle B\rangle \rangle A \\
\langle a\langle B\rangle \rangle A\rightarrow \langle a[B]\rangle A,\
where\ B\ is\ syntactically\ a\ valid\ process\ in\ the\ higher\  \\
\quad \quad \quad \quad order\ pi-calculus.
\end{array}
$$

Intuitively, axiom $a\circledR A\rightarrow ({\bf N}b)b\circledR A\{b/a\}$
means that if process $P$ satisfies $(\nu a)A$ and $b$ is a fresh name then $%
P$ satisfies $(\nu b)A\{b/a\}.$ Axiom $\overline{a}\langle B\rangle
.A\rightarrow \langle \overline{a}\langle B\rangle \rangle A$ means that an
output prefix process can perform an output action, which is a spatial
logical version of Rule $OUT$ in the labelled transition system of higher
order $\pi $-calculus. Axiom $(a\odot U.A\wedge ((\tilde \ominus
)B\leftrightarrow B))\rightarrow \langle a[B]\rangle A\{B/U\}$ means that an
input prefix process can perform an input action, which is a spatial logical
version of Rule $IN$ in the labelled transition system of higher order $\pi $%
-calculus. Axiom $(((\ominus b_1,...,\ominus b_n)B\leftrightarrow B)\wedge
((\tilde \ominus )C\leftrightarrow C))\rightarrow ((\langle \overline{a}%
\langle b_1\circledR ...b_n\circledR C\rangle \rangle A)|\langle a[C]\rangle
B\linebreak \rightarrow \langle \tau \rangle b_1\circledR ...b_n\circledR %
(A|B))$ is a spatial logical version of Rule $COM$. Other axioms and rules
are spatial logical version of structural congruence rules or labelled
transition rules similarly.

\subsection{Soundness of $SL$}

Inference system of $SL$ is said to be sound with respect to processes if
every formula provable in $SL$ is valid with respect to processes.

Now, we can prove the soundness of inference system $S$ of $SL$:

{\bf Proposition 1} $\Gamma \vdash _SA\Rightarrow \Gamma \models _{SL}A$

$Proof.$ See Appendix A.\TeXButton{End Proof}{\endproof}

\subsection{Incompleteness of $SL$}

The system $SL$ is complete with respect to processes if every formula valid
with respect to processes is provable in $SL$. For a logic, completeness is
an important property. The soundness and completeness provide a tight
connection between the syntactic notion of provability and the semantic
notion of validity. Unfortunately, by the compactness property \cite{Cha77},
the inference system of $SL$ is not complete.

The depth of higher order processes in $Pr$, is defined as below:

{\bf Definition} {\bf 10} $d(0)=0$; $d(U)=0$; $d(a(U).P)=1+d(P)$; $d(%
\overline{a}\langle E\rangle .P)=1+d(E)+d(P)$; $d(P_1|P_2)=d(P_1)+d(P_2);$ $%
d((\nu a)P)=d(P)$.

{\bf Lemma 1} For any $P\in Pr,$ there exists $n,$ such that $d(P)=n.$

$Proof.$ Induction on the structure of $P$.

{\bf Proposition 2 }There is no finite sound inference system $AX$ such that
$\Gamma \models _{SL}A\Rightarrow \Gamma \vdash _{AX}A.$

$Proof.$ See Appendix B.\TeXButton{End Proof}{\endproof}

\subsection{Spatial Logic as a Specification of Processes}

In the refinement calculus \cite{MGRV94}, imperative programming languages
are extended by specification statements, which specify parts of a program
``yet to be developed''. Then the development of a program begins with a
specification statement, and ends with an executable program by refining a
specification to its possible implementations. In this paper, we generalize
this idea to the case of process calculi. Roughly speaking, we extend
processes to spatial logic formulas which are regarded as the specification
statements. Processes can be regarded as a special kind of spatial logic.
One can view the intensional operators of spatial logic as the ``executable
program statements'', for example, $\overline{a}\langle P\rangle .Q,$ $P|Q$
and etc; and view the extensional operators of spatial logic as the
``specification statements'', for example, $A\triangleright B,$ $A\setminus
\overline{b}$ and etc. For example, $(b\odot Y.\overline{a}\langle Y\rangle
.A_1\triangleright \langle \tau \rangle A_2)\setminus \overline{b}|(d\odot Y.%
\overline{c}\langle B_1\rangle .Y\triangleright \langle \tau \rangle
B_2)\setminus \overline{d}$ represents a specification statement which
describes a process consisting of a parallel of two processes satisfying
statements $(b\odot Y.\overline{a}\langle Y\rangle .A_1\triangleright
\langle \tau \rangle A_2)\setminus \overline{b}$ and $(d\odot Y.\overline{c}%
\langle B_1\rangle .Y\triangleright \langle \tau \rangle B_2)\setminus
\overline{d}$ respectively. Furthermore, $(b\odot Y.\overline{a}\langle
Y\rangle .A_1\triangleright \langle \tau \rangle A_2)\setminus \overline{b}$
represents a specification which describes a process $P$ such that $%
\overline{a}\langle P\rangle .Q$ satisfies $A_2$ for any $Q$ satisfying $A_1.
$ Similarly, $(d\odot Y.\overline{c}\langle B_1\rangle .Y\triangleright
\langle \tau \rangle B_2)\setminus \overline{d}$ represents a specification
statement which describes a process $M$ such that $\overline{c}\langle
N\rangle .M$ satisfying $B_2$ for any $N$ satisfying $B_1.$ We can also
define refinement relation on spatial logic formulas. Intuitively, if $%
\models _{SL}A\rightarrow B,$ then $A$ refines $B.$ For example, $a\circledR %
(a\odot X.d.X|\overline{a}\langle c.0\rangle .e.0)$ refines $a\circledR %
(\langle a[c.0]\rangle d.c.0|\langle \overline{a}\langle c.0\rangle \rangle
e.0).$ Based on spatial logic, one may develop a theory of refinement for
concurrent processes. This will be a future research direction for us.

\subsection{Processes as Special Formulas of Spatial Logic}

Any process can be regarded as a special formula of spatial logic. For
example, (${\bf N}a)a\circledR ({\bf N}X)(a\odot X.d.X|\overline{a}\langle
c.0\rangle .e.0)$ is a spatial logic formula, which represents the process
which is structural congruent to $(\nu a)(a(X).d.X|\overline{a}\langle
c.0\rangle .e.0).$ Furthermore, in this section, we will show that
structural congruence and labelled transition relation can be reformulated
as the logical relation of spatial logical formulas.

{\bf Definition 11 }The translating function $T^{PS}$ is defined inductively
as follows:

$T^{PS}(P)\stackrel{def}{=}P$ for process $P$ that has no operators of $(\nu
a)\cdot $, or $a(X).\cdot ;$

$T^{PS}((\nu a)P)\stackrel{def}{=}({\bf H}a)T^{PS}(P);$

$T^{PS}(a(X).P)\stackrel{def}{=}(a{\bf H}X)T^{PS}(P).$

{\bf Proposition 3 }For any $P,Q\in Pr^c,$ $P\equiv Q$ $\Leftrightarrow
P\models _{SL}T^{PS}(Q)$ and $Q\models _{SL}T^{PS}(P)\Leftrightarrow
T^{PS}(P)\vdash _{SL}T^{PS}(Q)$ and $T^{PS}(Q)\vdash _{SL}T^{PS}(P).$

$Proof.$ See Appendix C.\TeXButton{End Proof}{\endproof}

{\bf Proposition 4 }For any $P,Q\in Pr^c,$ $P\stackrel{\alpha }{%
\longrightarrow }Q$ $\Leftrightarrow P\models _{SL}\langle \alpha \rangle
T^{PS}(Q)\Leftrightarrow T^{PS}(P)\vdash _{SL}\langle \alpha \rangle
T^{PS}(Q).$

$Proof.$ See Appendix D.\TeXButton{End Proof}{\endproof}

Although Proposition 2 states that the inference system is not complete,
Propositions 3 and 4 show that this inference system is complete with
respect to structural congruence and labelled transition relation of
processes.

\subsection{Behavioral Equivalence Relation of Spatial Logic}

In \cite{Cao06a}, we introduced a spatial logic called $L$, and proved that $%
L$ gives a characterization of context bisimulation.

{\bf Definition 12 }\cite{Cao06a} Syntax of logic $L$

$A::=\neg A$ \TEXTsymbol{\vert} $A_1\wedge A_2$ \TEXTsymbol{\vert} $\langle
a\langle \top \rangle \rangle \top $ \TEXTsymbol{\vert} $\langle \overline{a}%
\langle \top \rangle \rangle \top $ \TEXTsymbol{\vert} $\langle \tau \rangle
A$ \TEXTsymbol{\vert} $A_1\triangleright A_2$.

It is easy to see that $L$ is a sublogic of $SL$.

In \cite{Cao06a}, we proved the equivalence between $\sim _{Ct}$ and logical
equivalence with respect to $L.$

{\bf Proposition 5 }\cite{Cao06a}{\bf \ }For any $P,Q\in Pr^c,$ $P\sim
_{Ct}Q\Leftrightarrow $for any formula $A\in L$, $P\models _LA$ iff $%
Q\models _LA.$

{\bf Definition 13} $A$ and $B$ are behavioral equivalent with respect to $L$%
, written $A\sim _LB,$ iff for any formula $C\in L$, $\models
_{SL}A\rightarrow C$ iff $\models _{SL}B\rightarrow C.$

By Proposition 5, it is easy to get the following corollary, which
characterize $\sim _{Ct}\ $by $SL$ property.

{\bf Corollary 1} For any $P,Q\in Pr^c,$ $P\sim _{Ct}Q\Leftrightarrow P\sim
_LQ.$

Relation $\sim _L$ is a binary relation on spatial logical formulas. The
above results show that $\sim _L$ gives a logical characterization of
bisimulation when formulas are in the form of processes. Moreover, relation $%
\sim _L$ also gives a possibility to generialize bisimulation on processes
to that on spatial logical formulas. Since we have discussed that spatial
logical formulas can be regarded as specifications of processes, we may get
a concept of bisimulation on specifications of processes based on $\sim _L.$

\section{Logics for Weak Semantics}

In this section, we present a logic for weak semantics, named $WL$. Roughly
speaking, in this logic, action temporal operators $\langle \tau \rangle ,$ $%
\langle a\langle A\rangle \rangle ,$ $\langle a[A]\rangle $ and $\langle
\overline{a}\langle A\rangle \rangle $ in $SL$ are replaced by the weak
semantics version of operators $\langle \langle \varepsilon \rangle \rangle
, $ $\langle \langle a\langle A\rangle \rangle \rangle ,$ $\langle \langle
a[A]\rangle \rangle $ and $\langle \langle \overline{a}\langle A\rangle
\rangle \rangle .$ Almost all definitions and results of $SL$ can be
generalized to $WL$.

\subsection{Syntax and Semantics of Logic $WL$}

Now we introduce a logic called $WL,$ which is a weak semantics version of
spatial logic.

{\bf Definition 14} Syntax of logic $WL$

$A::=\top |$ $\bot |$ $\neg A$ \TEXTsymbol{\vert} $A_1\wedge A_2$
\TEXTsymbol{\vert} $\langle \langle \varepsilon \rangle \rangle A$
\TEXTsymbol{\vert} $\langle \langle a\langle A_1\rangle \rangle \rangle A_2$
\TEXTsymbol{\vert} $\langle \langle a[A_1]\rangle \rangle A_2$ \TEXTsymbol{%
\vert} $\langle \langle \overline{a}\langle A_1\rangle \rangle \rangle A_2$
\TEXTsymbol{\vert} $0$ $|$ $X$ $|$ $a\odot X.A$ \TEXTsymbol{\vert} $%
A\setminus a\odot X$ \TEXTsymbol{\vert} $\overline{a}\langle A_1\rangle .A_2$
\TEXTsymbol{\vert} $A\setminus \overline{a}$ \TEXTsymbol{\vert} $A_1|A_2$
\TEXTsymbol{\vert} $A_1\triangleright A_2$ \TEXTsymbol{\vert} $a\circledR A$
\TEXTsymbol{\vert} $A\oslash a$ \TEXTsymbol{\vert} $({\bf N}x)A$ $|$ $({\bf N%
}X)A$ $|$ $(\ominus a)A$ \TEXTsymbol{\vert} $(\tilde \ominus )A$ \TEXTsymbol{%
\vert} $a\not =b$

{\bf Definition 15} Semantics of logic $WL$

Semantics of formulas of $WL$ can be the same as formulas of $SL$, except
that semantics of operators $\langle \langle \varepsilon \rangle \rangle ,$ $%
\langle \langle a\langle A\rangle \rangle \rangle ,$ $\langle \langle
a[A]\rangle \rangle $ and $\langle \langle \overline{a}\langle A\rangle
\rangle \rangle $ should be defined as follows:

$[[\langle \langle \varepsilon \rangle \rangle A]]_{Pr}=\{P$ $|$ $\exists Q$%
. $P\stackrel{\varepsilon }{\Longrightarrow }Q$ and $Q\in [[A]]_{Pr}\}$

$[[\langle \langle a\langle A_1\rangle \rangle \rangle A_2]]_{Pr}=\{P$ $|$ $%
\exists P_1,P_2$. $P\stackrel{a\langle P_1\rangle }{\Longrightarrow }P_2,$ $%
P_1\in [[A_1]]_{Pr}$ and $P_2\in [[A_2]]_{Pr}\}$

$[[\langle \langle a[A_1]\rangle \rangle A_2]]_{Pr}=\{P$ $|$ $\forall R,R\in
[[A_1]]_{Pr},\exists Q$. $P\stackrel{a\langle R\rangle }{\Longrightarrow }Q$
and $Q\in [[A_2]]_{Pr}\}$

$[[\langle \langle \overline{a}\langle A_1\rangle \rangle \rangle
A_2]]_{Pr}=\{P$ $|$ $\exists P_1,P_2$. $P\stackrel{(\nu \widetilde{b})%
\overline{a}\langle P_1\rangle }{\Longrightarrow }P_2,$ $(\nu \widetilde{b}%
)P_1\in [[A_1]]_{Pr}$ and $P_2\in [[A_2]]_{Pr}\}$

\subsection{Inference System of $WL$}

The inference system of $WL$ is similar to the inference system of $SL$
except that any inference rule about action temporal operators $\langle \tau
\rangle ,$ $\langle a\langle A\rangle \rangle ,$ $\langle a[A]\rangle $ and $%
\langle \overline{a}\langle A\rangle \rangle $ in $SL$ is replaced by one of
the following inference rules.

$\langle \langle \alpha \rangle \rangle \bot \rightarrow \bot $

$\langle \langle \alpha \rangle \rangle A,A\rightarrow B\vdash \langle
\langle \alpha \rangle \rangle B$

$\langle \langle \alpha \rangle \rangle A,A\rightarrow \langle \langle
\varepsilon \rangle \rangle B\vdash \langle \langle \alpha \rangle \rangle B$

$\langle \langle \varepsilon \rangle \rangle A,A\rightarrow \langle \langle
\alpha \rangle \rangle B\vdash \langle \langle \alpha \rangle \rangle B$

$\langle \langle \overline{a}\langle B\rangle \rangle \rangle A,C\rightarrow
B\vdash \langle \langle \overline{a}\langle C\rangle \rangle \rangle A\quad $

$\langle \langle a[B]\rangle \rangle A,C\rightarrow B\vdash \langle \langle
a[C]\rangle \rangle A\quad $

$\overline{a}\langle B\rangle .A\rightarrow \langle \langle \overline{a}%
\langle B\rangle \rangle \rangle A$

$(a\odot U.A\wedge ((\tilde \ominus )B\leftrightarrow B))\rightarrow \langle
\langle a[B]\rangle \rangle A\{B/U\}$

$(\langle \langle \varepsilon \rangle \rangle A)|B\rightarrow \langle
\langle \varepsilon \rangle \rangle (A|B)$

$(\langle \langle a\langle C\rangle \rangle \rangle A)|B\rightarrow \langle
\langle a\langle C\rangle \rangle \rangle (A|B)$

$(((\ominus b_1,...,\ominus b_n)B\leftrightarrow B)\wedge ((\tilde \ominus
)C\leftrightarrow C))\rightarrow $

$\quad \quad \quad \quad ((\langle \langle \overline{a}\langle b_1\circledR %
...b_n\circledR C\rangle \rangle \rangle A)|B\rightarrow \langle \langle
\overline{a}\langle b_1\circledR ...b_n\circledR C\rangle \rangle \rangle
(A|B))$

$(((\ominus b_1,...,\ominus b_n)B\leftrightarrow B)\wedge ((\tilde \ominus
)C\leftrightarrow C))\rightarrow $

$\quad \quad \quad \quad ((\langle \langle \overline{a}\langle b_1\circledR %
...b_n\circledR C\rangle \rangle \rangle A)|\langle \langle a[C]\rangle
\rangle B\rightarrow \langle \langle \varepsilon \rangle \rangle b_1%
\circledR ...b_n\circledR (A|B))$

$a\circledR \langle \langle \varepsilon \rangle \rangle A\rightarrow \langle
\langle \varepsilon \rangle \rangle a\circledR A$

$(a\neq b\wedge (((\ominus a)B\wedge (\tilde \ominus )B)\leftrightarrow
B))\rightarrow (a\circledR \langle \langle b\langle B\rangle \rangle \rangle
A\rightarrow \langle \langle b\langle B\rangle \rangle \rangle a\circledR A)$

$(\wedge _{i=1}^na\neq b_i\wedge a\neq c\wedge ((\ominus a)B\leftrightarrow
B)\wedge ((\tilde \ominus )B\leftrightarrow B))\rightarrow $

$\quad \quad \quad \quad (a\circledR \langle \langle \overline{c}\langle b_1%
\circledR ...b_n\circledR B\rangle \rangle \rangle A\rightarrow \langle
\langle \overline{c}\langle b_1\circledR ...b_n\circledR B\rangle \rangle
\rangle a\circledR A)$

$(a\neq b\wedge \wedge _{i=1}^nb\neq c_i\wedge \ (B\rightarrow \neg (\ominus
b)\top )\wedge ((\tilde \ominus )B\leftrightarrow B))\rightarrow $

$\quad \quad \quad \quad (b\circledR \langle \langle \overline{a}\langle c_1%
\circledR ...c_n\circledR B\rangle \rangle \rangle A\rightarrow \langle
\langle \overline{a}\langle b\circledR c_1\circledR ...c_n\circledR B\rangle
\rangle \rangle A)$

$\langle \langle a[B]\rangle \rangle A\rightarrow \langle \langle a\langle
B\rangle \rangle \rangle A$

$\langle \langle a\langle B\rangle \rangle \rangle A\rightarrow \langle
\langle a[B]\rangle \rangle A,$ $where\ B\ is\ syntactically$ $a$ $valid$ $%
process$ $in$ $the$ $higher$ $order$ $pi-calculus$.

The above axioms and rules are weak semantics version of corresponding
axioms and rules in $SL.$ We name the above inference system of $WL$ as $W$.

The soundness and incompleteness of inference system $W$ of $WL$ can be
given similarly as the case of $SL$:

{\bf Proposition 6} $\Gamma \vdash _WA\Rightarrow \Gamma \models _{WL}A$

{\bf Proposition 7 }There is no finite sound inference system $AX$ such that
$\Gamma \models _{WL}A\Rightarrow \Gamma \vdash _{AX}A.$

Similar to Proposition 4, we show that many-steps transition relation is
provable in $WL.$

{\bf Proposition 8 }For any $P,Q\in Pr^c,$ $P\stackrel{\alpha }{%
\Longrightarrow }Q$ $\Leftrightarrow P\models _{WL}\langle \langle \alpha
\rangle \rangle T^{PS}(Q)\Leftrightarrow T^{PS}(P)\vdash _{WL}\langle
\langle \alpha \rangle \rangle T^{PS}(Q).$

Since structural congruence and labelled transition relation are central
concepts in the theory of processes, and they can be characterized in $WL$,
the above propositions give a possible approach to reduce the theory of
processes to the theory of spatial logic in the case of weak semantics.

\section{Adding $\mu $-Operator to $SL$}

In this section, we add $\mu $-operator \cite{AN01} to $SL.$ We refer to
this new logic as $\mu SL.$ We will show that $WL$ is a sublogic of $\mu SL.$

\subsection{Syntax and Semantics of $\mu SL$}

The formula of $\mu SL$ is the same as the formula of $SL$ except that the
following $\mu $-calculus formula is added:

If $A(X)\in \mu SL$, then $\mu X.A(X)\in \mu SL,$ here $X$ occurs positively
in $A(X)$, i.e., all free occurrences of $X$ fall under an even number of
negations.$.$

The model of $\mu SL$ is the same as $SL$. We write such set of processes in
which $A$ is true as $[[A]]_{Pr}^e,$ where $e$: $Var\rightarrow 2^{Pr}$ is
an environment. We denote by $e[X\leftarrow W]$ a new environment that is
the same as $e$ except that $e[X\leftarrow W](X)=W.$ The set $[[A]]_S^e$ is
the set of processes that satisfy $A$. In the following, we abbreviate $A(B)$
as $A\{B/X\},\ $and abbreviate $A^{n+1}(B)$ as $A(A^n(B))$ where $A^0(B)$ is
$B.$

Semantics of $\mu $-operator is given as following:

$[[\mu X.A(X)]]_{Pr}^e=\cap \{W\subseteq Pr$ $|$ $[[A(X)]]_{Pr}^{e[X%
\leftarrow W]}\subseteq W\}.$

In $\mu $-calculus \cite{AN01}, it is well known that $[[\mu
X.A(X)]]_{Pr}^e=[[A^1(\bot )]]_{Pr}^e\cup [[A^2(\bot )]]_{Pr}^e\linebreak %
\cup ...$

\subsection{Inference System of $\mu SL$}

Inference system of $\mu SL$ is the combination of the following two rules
of $\mu $-calculus \cite{AN01} and the inference system of $SL.$

$A(\mu X.A(X))\rightarrow \mu X.A(X)$

$\dfrac{A(B)\rightarrow B}{\mu X.A(X)\rightarrow B}$

We name the above inference system of $\mu SL$ as $M$.

The soundness and incompleteness of inference system $M$ of $\mu SL$ can be
given as in the case of $SL.$

{\bf Proposition 9} $\Gamma \vdash _MA\Rightarrow \Gamma \models _{\mu SL}A$

{\bf Proposition 10 }There is no finite sound inference system $AX$ such
that $\Gamma \models _{\mu SL}A\Rightarrow \Gamma \vdash _{AX}A.$

\subsection{Expressivity of $\mu SL$}

In this section, we will discuss the expressive power of $\mu SL.$ We will
prove that $WL$ is a sublogic of $\mu SL$ and give a function which can
translates a $WL$ formula into an equivalent $\mu SL$ formula.

Now we can give a translating function from $WL$ formula to $\mu SL$ formula:

{\bf Definition 16 }The translating function $T$ is defined inductively as
follows:

$T^{WM}(A)\stackrel{def}{=}A$ for proposition $A$ of $WL$ that is not in the
form of $\langle \langle \varepsilon \rangle \rangle A$, $\langle \langle
a\langle A_1\rangle \rangle \rangle A_2$, $\langle \langle a[A_1]\rangle
\rangle A_2$ or $\langle \langle \overline{a}\langle A_1\rangle \rangle
\rangle A_2.$

$T^{WM}(\langle \langle \varepsilon \rangle \rangle A)\stackrel{def}{=}\mu
X.(T^{WM}(A)\vee \langle \tau \rangle X)$

$T^{WM}(\langle \langle a\langle A_1\rangle \rangle \rangle A_2)\stackrel{def%
}{=}\mu X.(\langle a\langle T^{WM}(A_1)\rangle \rangle (\mu
Y.(T^{WM}(A_2)\vee \langle \tau \rangle Y))\vee \langle \tau \rangle X)$

$T^{WM}(\langle \langle a[A_1]\rangle \rangle A_2)\stackrel{def}{=}\mu
X.(\langle a[T^{WM}(A_1)]\rangle (\mu Y.(T^{WM}(A_2)\vee \langle \tau
\rangle Y)\vee \langle \tau \rangle X)$

$T^{WM}(\langle \langle \overline{a}\langle A_1\rangle \rangle \rangle A_2)%
\stackrel{def}{=}\mu X.(\langle \overline{a}\langle T^{WM}(A_1)\rangle
\rangle (\mu Y.(T^{WM}(A_2)\vee \langle \tau \rangle Y)\vee \langle \tau
\rangle X)$

The following proposition states the correctness of translating function $%
T^{WM}.$

{\bf Proposition 11 }For any $A\in WL,$ $T^{WM}(A)\in \mu SL;$ for any $P\in
Pr,$ $P\models _{\mu SL}T^{WM}(A)\Leftrightarrow P\models _{WL}A.$

$Proof:$ See Appendix E. \TeXButton{End Proof}{\endproof}

In $\mu SL,$ we can also define the replication operator:

{\bf Definition 17 }$!A\stackrel{def}{=}\neg \mu X.\neg (A|\neg X)$

{\bf Proposition} {\bf 12 }$\vdash _{\mu SL}A|!A\leftrightarrow !A$

$Proof:$ See Appendix F. \TeXButton{End Proof}{\endproof}

The above results show that $WL$ is a sublogic of $\mu SL.$ Therefore $\mu SL
$ can be used as a uniform logic framework to study both the strong
semantics and the weak semantics of higher order $\pi $-calculus.

\section{Conclusions}

Spatial logic was proposed to describe structural and behavioral properties
of processes. There are many papers on spatial logic and process calculi.
Spatial logic is related to some topics on process calculi, such as model
checking, structural congruence, bisimulation and type system. In \cite{CG00}%
, a spatial logic for ambients calculus was studied, and a model checking
algorithm was proposed. Some axioms of spatial logic were given, but the
completeness of logic was not studied. Most spatial logics for concurrency
are intensional \cite{San01}, in the sense that they induce an equivalence
that coincides with structural congruence, which is much finer than
bisimilarity. In \cite{Hir04}, Hirschkoff studied an extensional spatial
logic. This logic only has spatial composition adjunct ($\triangleright $),
revelation adjunct ($\oslash $), a simple temporal modality ($\langle
\rangle $), and an operator for fresh name quantification. For $\pi $%
-calculus, this extensional spatial logic was proven to induce the same
separative power as strong early bisimilarity. In \cite{Cao06a}, context
bisimulation of higher order $\pi $-calculus was characterized by an
extensional spatial logic. In \cite{Cai08}, a type system of processes based
on spatial logic was given, where types are interpreted as formulas of
spatial logic.

In this paper, we want to show that the theory of processes can be reduced
to the theory of spatial logics. We firstly defined a logic $SL${\it , }%
which comprises some temporal operators and spatial operators. We gave the
inference system of $SL$ and showed the soundness and incompleteness of $SL.$
Furthermore, we showed that structural congruence and transition relation of
higher order $\pi $-calculus can be reduced to the logical relation of $SL$
formulas. We also showed that bisimulations in higher order $\pi $-calculus
can be characterized by a sublogic of $SL$. Furthermore, we propose a weak
semantics version of $SL$, called $WL$. At last, we add $\mu $-operator to $%
SL.$ The new logic is named $\mu SL.$ We studied the expressive power of
this extension. These results can be generalized to other process calculi.
Since some important concepts of processes can be described in spatial
logic, we think that this paper may give an approach of reducing the study
of processes to the study of spatial logic. The further work for us is to
develop a refinement calculus \cite{MGRV94} for concurrent processes based
on our spatial logic.

\begin{center}
{\bf Appendix A. Proof of Proposition 1}
\end{center}

{\bf Proposition 1} $\Gamma \vdash _{SL}A\Rightarrow \Gamma \models _{SL}A$

$Proof.$ It is enough by proving that every axiom and every inference rule
of inference system is sound. We only discuss the following cases:

Case (1): Axiom $a\circledR ((\ominus a)A|B)\leftrightarrow (\ominus a)A|a%
\circledR B.$

Suppose $P\in [[a\circledR ((\ominus a)A|B)]],$ then $P\equiv (\nu
a)(P_1|P_2)$, $a\notin fn(P_1),$ $P_1\in [[A]]$ and $P_2\in [[B]].$
Therefore we have $P\equiv (\nu a)(P_1|P_2)\equiv P_1|(\nu a)P_2,$ $P\in
[[(\ominus a)A|a\circledR B]].$ Hence $a\circledR ((\ominus
a)A|B)\leftrightarrow (\ominus a)A|a\circledR B.$ The inverse case is
similar.

Case (2): Axiom $a\neq b\rightarrow ((\ominus a)\overline{b}\langle B\rangle
.A\leftrightarrow \overline{b}\langle (\ominus a)B\rangle .(\ominus a)A).$

Suppose $a\neq b$ and $P\in [[(\ominus a)\overline{b}\langle B\rangle .A]],$
then $P\equiv \overline{b}\langle P_1\rangle .P_2$, $a\notin fn(P_1),$ $%
a\notin fn(P_2),$ $P_1\in [[B]]$ and $P_2\in [[A]].$ Therefore we have $%
P_1\in [[(\ominus a)B]]$ and $P_2\in [[(\ominus a)A]],$ $P\in [[\overline{b}%
\langle (\ominus a)B\rangle .(\ominus a)A)]].$ Hence $a\neq b\rightarrow
((\ominus a)\overline{b}\langle B\rangle .A\rightarrow \overline{b}\langle
(\ominus a)B\rangle .(\ominus a)A).$ The inverse case is similar.

Case (3): Axiom $(A|A\triangleright B)\rightarrow B.$

Suppose $P\in [[A|A\triangleright B]],$ then $P\equiv P_1|P_2,$ $P_1\in
[[A]] $ and $P_2\in [[A\triangleright B]].$ Therefore, $P\equiv P_1|P_2\in
[[A|A\triangleright B]].$ Hence $(A|A\triangleright B)\rightarrow B.$

Case (4): Axiom $A\rightarrow (B\triangleright A|B).$

Suppose $P\in [[A]],$ then for any $Q\in [[B]],$ $P|Q\in [[A|B]].$ Hence $%
A\rightarrow (B\triangleright A|B).$

Case (5): Axiom $(((\ominus b_1,...,\ominus b_n)B\leftrightarrow B)\wedge
((\tilde \ominus )C\leftrightarrow C))\rightarrow \linebreak ((\langle
\overline{a}\langle b_1\circledR ...b_n\circledR C\rangle \rangle
A)|B\rightarrow \langle \overline{a}\langle b_1\circledR ...b_n\circledR %
C\rangle \rangle (A|B)).$

Suppose $P\in [[(\langle \overline{a}\langle b_1\circledR ...b_n\circledR %
C\rangle \rangle A)|B]],$ then $P\equiv P_1|P_2,$ $P_1\stackrel{(\nu
b_1,...,b_n)\overline{a}\langle Q\rangle }{\longrightarrow }P_1^{\prime },$ $%
P_1^{\prime }\in [[A]]$, $P_2\in [[B]]$ and $Q\in [[C]].$ Since $(\ominus
b_1,...,b_n)B\leftrightarrow B,$ $\{b_1,...,b_n\}\cap fn(P_2)=\emptyset .$
Therefore we have $P_1|P_2\stackrel{(\nu b_1,...,b_n)\overline{a}\langle
Q\rangle }{\longrightarrow }P_1^{\prime }|P_2.$ Hence $(((\ominus
b_1,...,\ominus b_n)B\linebreak \leftrightarrow B)\wedge ((\tilde \ominus
)C\leftrightarrow C))\rightarrow ((\langle \overline{a}\langle b_1\circledR %
...b_n\circledR C\rangle \rangle A)|B\rightarrow \langle \overline{a}\langle
b_1\circledR ...b_n\circledR C\rangle \rangle \linebreak (A|B)).$

Case (6): Axiom $(((\ominus b_1,...,\ominus b_n)B\leftrightarrow B)\wedge
((\tilde \ominus )C\leftrightarrow C))\rightarrow \linebreak ((\langle
\overline{a}\langle b_1\circledR ...b_n\circledR C\rangle \rangle A)|\langle
a[C]\rangle B\rightarrow \langle \tau \rangle b_1\circledR ...b_n\circledR %
(A|B)).$

Suppose $P\in [[(\langle \overline{a}\langle b_1\circledR ...b_n\circledR %
C\rangle \rangle A)|\langle a[C]\rangle B]],$ then $P\equiv P_1|P_2,$ $%
\linebreak P_1\stackrel{(\nu b_1,...,b_n)\overline{a}\langle Q\rangle }{%
\longrightarrow }P_1^{\prime },$ $P_2\stackrel{a\langle Q\rangle }{%
\longrightarrow }P_2^{\prime },$ $P_1^{\prime }\in [[A]]$, $P_2^{\prime }\in
[[B]]$ and $Q\in [[C]].$ Since $(\ominus b_1,...,b_n)B\leftrightarrow B,$ $%
\{b_1,...,b_n\}\cap fn(P_2^{\prime })=\emptyset .$ Therefore we have $P_1|P_2%
\stackrel{\tau }{\longrightarrow }(\nu b_1,...,b_n)(P_1^{\prime
}|P_2^{\prime }).$ Hence $(((\ominus b_1,...,\ominus b_n)B\leftrightarrow
B)\wedge ((\tilde \ominus )C\leftrightarrow C))\rightarrow \linebreak %
((\langle \overline{a}\langle b_1\circledR ...b_n\circledR C\rangle \rangle
A)|\langle a[C]\rangle B\rightarrow \langle \tau \rangle b_1\circledR ...b_n%
\circledR (A|B)).$

Case (7): Axiom $(\wedge _{i=1}^na\neq b_i\wedge a\neq c\wedge ((\ominus
a)B\leftrightarrow B)\wedge ((\tilde \ominus )B\leftrightarrow
B))\rightarrow (a\circledR \langle \overline{c}\langle b_1\circledR ...b_n%
\circledR B\rangle \rangle A\rightarrow \langle \overline{c}\langle b_1%
\circledR ...b_n\circledR B\rangle \rangle a\circledR A).$

Suppose $P\in [[a\circledR \langle \overline{c}\langle b_1\circledR ...b_n%
\circledR B\rangle \rangle A]],$ then $P\equiv (\nu a)P_1,$ $P_1\stackrel{%
(\nu b_1,...,b_n)\overline{c}\langle Q\rangle }{\longrightarrow }P_1^{\prime
},$ $Q\in [[B]],$ $P_1^{\prime }\in [[A]].$ Since $\wedge _{i=1}^na\neq
b_i\wedge a\neq c\wedge ((\ominus a)B\leftrightarrow B)\wedge ((\tilde
\ominus )B\leftrightarrow B),$ $a\notin n(Q).$ Therefore we have $P\equiv
(\nu a)P_1\stackrel{(\nu b_1,...,b_n)\overline{c}\langle Q\rangle }{%
\longrightarrow }(\nu a)P_1^{\prime }.$ Hence $(\wedge _{i=1}^na\neq
b_i\wedge a\neq c\wedge ((\ominus a)B\leftrightarrow B)\wedge ((\tilde
\ominus )B\leftrightarrow B))\rightarrow (a\circledR \langle \overline{c}%
\langle b_1\circledR ...b_n\circledR B\rangle \rangle A\rightarrow \langle
\overline{c}\langle b_1\circledR ...b_n\circledR B\rangle \rangle a\circledR %
A).$

Case (8): Axiom $(a\neq b\wedge \wedge _{i=1}^nb\neq c_i\wedge (\
B\rightarrow \neg (\ominus b)\top )\wedge ((\tilde \ominus )B\leftrightarrow
B))\rightarrow (b\circledR \langle \overline{a}\langle c_1\circledR ...c_n%
\circledR B\rangle \rangle A\rightarrow \langle \overline{a}\langle b%
\circledR c_1\circledR ...c_n\circledR B\rangle \rangle A).$

Suppose $P\in [[b\circledR \langle \overline{a}\langle c_1\circledR ...c_n%
\circledR B\rangle \rangle A]],$ then $P\equiv (\nu b)P_1,$ $P_1\stackrel{%
(\nu c_1,...,c_n)\overline{a}\langle Q\rangle }{\longrightarrow }P_1^{\prime
},$ $Q\in [[B]],$ $P_1^{\prime }\in [[A]].$ Since $a\neq b\wedge \wedge
_{i=1}^nb\neq c_i\wedge (\ B\rightarrow \neg (\ominus b)\top )\wedge
((\tilde \ominus )B\leftrightarrow B),$ $b\in fn(Q).$ Therefore we have $%
P\equiv (\nu b)P_1\stackrel{(\nu b)(\nu c_1,...,c_n)\overline{a}\langle
Q\rangle }{\longrightarrow }P_1^{\prime }.$ Hence $(a\neq b\wedge \wedge
_{i=1}^nb\neq c_i\wedge (B\rightarrow \neg (\ominus b)\top )\wedge ((\tilde
\ominus )B\leftrightarrow B))\rightarrow (b\circledR \langle \overline{a}%
\langle c_1\circledR ...c_n\circledR B\rangle \rangle A\rightarrow \langle
\overline{a}\langle b\circledR c_1\circledR ...c_n\circledR B\rangle \rangle
A).$\TeXButton{End Proof}{\endproof}

\begin{center}
{\bf Appendix B. Proof of Proposition 2}
\end{center}

{\bf Proposition 2 }There is no finite sound inference system $AX$ such that
$\Gamma \models _{SL}A\Rightarrow \Gamma \vdash _{AX}A.$

$Proof.$ Let $\Phi =\{\overline{a}\langle 0\rangle .\top ,$ $\overline{a}%
\langle 0\rangle .\overline{a}\langle b.0\rangle .\top ,$ $\overline{a}%
\langle 0\rangle .\overline{a}\langle b.0\rangle .\overline{a}\langle
b.b.0\rangle .\top ,$ $\overline{a}\langle 0\rangle .\overline{a}\langle
b.0\rangle .\linebreak \overline{a}\langle b.b.0\rangle .\overline{a}\langle
b.b.b.0\rangle .\top ,...\}.$ It is easy to see that any finite subset of $%
\Phi $ can be satisfied in $Pr,$ but $\Phi $ can not be satisfied in $Pr.$
Suppose it is not true, let $P$ satisfies $\Phi .$ By Lemma 1, there exists $%
n,$ such that $d(P)=n.$ But for any $n$, there exists $\varphi _n$ in $\Phi $
such that for any $P$ satisfying $\varphi _n,$ $d(P)>n.$ This contradicts
the assumption. Therefore $\Phi $ can not be satisfied in $Pr.$

Suppose there is a finite inference system such that $\Gamma \models
_{SL}A\Rightarrow \Gamma \vdash _{SL}A.$ Since $\Phi $ can not be be
satisfied in $Pr,$ we have $\Phi \models _{SL}\bot .$ By the assumption, $%
\Phi \vdash _{SL}\bot .$ Hence there is a proof from $\Phi $ to $\bot $ in $%
SL.$ Since proof is a finite formula sequence, there is finite many formulas
$\varphi _i$ in $\Phi $ occur in the proof. Therefore we have $\wedge \Phi
_i\vdash _{SL}\bot ,$ where $\Phi _i=\{\varphi _i$ \TEXTsymbol{\vert} $%
\varphi _i$ is in the proof$\}.$ Then by the soundness of inference system
of $SL,$ we have that $\Phi _i$ is not satisfiable. Since $\Phi _i$ is a
finite subset of $\Phi ,$ this contradicts the assumption. Therefore $SL$
have no finite complete inference system.\TeXButton{End Proof}{\endproof}

\begin{center}
{\bf Appendix C. Proof of Proposition 3}
\end{center}

{\bf Proposition 3 }For any $P,Q\in Pr^c,$ $P\equiv Q$ $\Leftrightarrow
P\models _{SL}T^{PS}(Q)$ and $Q\models _{SL}T^{PS}(P)\Leftrightarrow
T^{PS}(P)\vdash _{SL}T^{PS}(Q)$ and $T^{PS}(Q)\vdash _{SL}T^{PS}(P).$

$Proof.$ It is trivial by the definition that $P\equiv Q$ $\Leftrightarrow
P\models _{SL}T^{PS}(Q)$ and $Q\models _{SL}T^{PS}(P).$ By the soundness, $%
T^{PS}(P)\vdash _{SL}T^{PS}(Q)\Rightarrow P\models _{SL}T^{PS}(Q).$ We only
need to prove $P\equiv Q$ $\Rightarrow T^{PS}(P)\vdash _{SL}T^{PS}(Q)$ and $%
T^{PS}(Q)\vdash _{SL}T^{PS}(P).$

We only discuss the following cases, other cases are similar or trivial:

Case (1): $(\nu m)(\nu n)P\equiv (\nu n)(\nu m)P:$ Since $m\circledR n%
\circledR T^{PS}(P)\leftrightarrow n\circledR m\circledR T^{PS}(P)$, we have$%
\ m\circledR n\circledR T^{PS}(P)\vdash _{SL}n\circledR m\circledR %
T^{PS}(P). $ The inverse case is similar.

Case (2): $(\nu a)(P|Q)\equiv P|(\nu a)Q\ $if $a\notin fn(P):$ Since $%
a\notin fn(P),$ $(\ominus a)T^{PS}(P)\leftrightarrow T^{PS}(P).$
Furthermore, since $a\circledR ((\ominus
a)T^{PS}(P)|T^{PS}(Q))\leftrightarrow (\ominus a)T^{PS}(P)|a\circledR %
T^{PS}(Q)$, we have $a\circledR (T^{PS}(P)|T^{PS}(Q))\vdash _{SL}T^{PS}(P)|a%
\circledR T^{PS}(Q).$ The inverse case is similar.\TeXButton{End Proof}
{\endproof}

\begin{center}
{\bf Appendix D. Proof of Proposition 4}
\end{center}

{\bf Proposition 4 }For any $P,Q\in Pr^c,$ $P\stackrel{\alpha }{%
\longrightarrow }Q$ $\Leftrightarrow P\models _{SL}\langle \alpha \rangle
T^{PS}(Q)\Leftrightarrow T^{PS}(P)\vdash _{SL}\langle \alpha \rangle
T^{PS}(Q).$

$Proof.$ It is trivial by the definition that $P\stackrel{\alpha }{%
\longrightarrow }Q$ $\Leftrightarrow P\models _{SL}\langle \alpha \rangle
T^{PS}(Q).$ By the soundness, $T^{PS}(P)\vdash _{SL}\langle \alpha \rangle
T^{PS}(Q)\Rightarrow P\models _{SL}\langle \alpha \rangle T^{PS}(Q).$ We
only need to prove $P\stackrel{\alpha }{\longrightarrow }Q$ $\Rightarrow
P\vdash _{SL}\langle \alpha \rangle T^{PS}(P).$

We apply the induction on the length of the inference tree of $P\stackrel{%
\alpha }{\longrightarrow }Q:$

Case (1): if the length is 0, then $P\stackrel{\alpha }{\longrightarrow }Q$
is in the form of $\overline{a}\langle E\rangle .K\stackrel{\overline{a}%
\langle E\rangle }{\longrightarrow }K$ or $a(U).K\stackrel{a\langle E\rangle
}{\longrightarrow }K\{E/U\}.$

Subcase (a): $\overline{a}\langle E\rangle .K\stackrel{\overline{a}\langle
E\rangle }{\longrightarrow }K:$ Since $\overline{a}\langle E\rangle
.T^{PS}(K)\rightarrow \langle \overline{a}\langle E\rangle \rangle
T^{PS}(K), $ we have $\overline{a}\langle E\rangle .T^{PS}(K)\vdash
_{SL}\langle \overline{a}\langle E\rangle \rangle T^{PS}(K).$

Subcase (b): $a(U).K\stackrel{a\langle E\rangle }{\longrightarrow }K\{E/U\}:$
Since $(a(U).T^{PS}(K)\wedge ((\tilde \ominus )T^{PS}(E)\leftrightarrow
T^{PS}(E)))\rightarrow \langle a[T^{PS}(E)]\rangle T^{PS}(K)\{T^{PS}(E)/U\},$
we have $a(U).T^{PS}(K)\vdash _{SL}\linebreak \langle a[T^{PS}(E)]\rangle
T^{PS}(K)\{T^{PS}(E)/U\}.$

Case (2): Assume the claim holds if length is $n$, now we discuss the case
that length is $n+1.$

Subcase (a): $\dfrac{M\stackrel{(\nu \widetilde{b})\overline{a}\langle
E\rangle }{\longrightarrow }M^{\prime }\quad N\stackrel{a\langle E\rangle }{%
\longrightarrow }N^{\prime }}{M|N\stackrel{\tau }{\longrightarrow }(\nu
\widetilde{b})(M^{\prime }|N^{\prime })}\widetilde{b}\cap fn(N)=\emptyset .$

Since $M\stackrel{(\nu \widetilde{b})\overline{a}\langle E\rangle }{%
\longrightarrow }M^{\prime },$ $N\stackrel{a\langle E\rangle }{%
\longrightarrow }N^{\prime },$ and $\widetilde{b}\cap fn(N)=\emptyset ,$ we
have $T^{PS}(M)\rightarrow \langle \overline{a}\langle \widetilde{b}%
\circledR T^{PS}(E)\rangle \rangle T^{PS}(M^{\prime })$, $%
T^{PS}(N)\rightarrow \langle a[T^{PS}(E)]\rangle T^{PS}(N^{\prime })$ and $%
(\ominus b_1,...,b_n)T^{PS}(E)\linebreak \leftrightarrow T^{PS}(E).$ By the
axiom: $(((\ominus b_1,...,b_n)T^{PS}(N)\leftrightarrow T^{PS}(N))\wedge
(\tilde \ominus )T^{PS}(E))\rightarrow ((\langle \overline{a}\langle b_1%
\circledR ...b_n\circledR T^{PS}(E)\rangle \rangle T^{PS}(M))|\langle
a[T^{PS}(E)]\rangle T^{PS}(N)\rightarrow \langle \tau \rangle b_1\circledR %
...b_n\circledR (T^{PS}(M)\linebreak |T^{PS}(N))),$ we have $P\equiv
T^{PS}(M)|T^{PS}(N)\vdash _{SL}\langle \tau \rangle b_1\circledR ...b_n%
\circledR (T^{PS}(M^{\prime })|T^{PS}(N^{\prime })).$

Subcase (b): $\dfrac{M\stackrel{b\langle E\rangle }{\longrightarrow }%
M^{\prime }}{(\nu a)M\stackrel{b\langle E\rangle }{\longrightarrow }(\nu
a)M^{\prime }}a\notin n(\alpha ).$

Since $M\stackrel{b\langle E\rangle }{\longrightarrow }M^{\prime }$ and $%
a\notin n(b\langle E\rangle ),$ we have $T^{PS}(M)\rightarrow \langle
b\langle T^{PS}(E)\rangle \rangle T^{PS}(M^{\prime })$ and $((\ominus
a)T^{PS}(E)\wedge (\tilde \ominus )T^{PS}(E))\leftrightarrow T^{PS}(E).$ By
the axiom $(a\neq b\wedge ((\ominus a)T^{PS}(E)\wedge (\tilde \ominus
)T^{PS}(E))\leftrightarrow T^{PS}(E))\rightarrow (a\circledR \langle
b\langle T^{PS}(E)\rangle \rangle T^{PS}(M)\rightarrow \langle b\langle
T^{PS}(E)\rangle \rangle a\circledR T^{PS}(M)),$ we have $T^{PS}(P)=a%
\circledR T^{PS}(M)\vdash _{SL}a\circledR \langle b\langle T^{PS}(E)\rangle
\rangle T^{PS}(M)\vdash _{SL}\linebreak \langle b\langle T^{PS}(E)\rangle
\rangle a\circledR T^{PS}(M)).$

Subcase (c): $\dfrac{M\stackrel{(\nu \widetilde{c})\overline{a}\langle
E\rangle }{\longrightarrow }M^{\prime }}{(\nu b)M\stackrel{(\nu b,\widetilde{%
c})\overline{a}\langle E\rangle }{\longrightarrow }M^{\prime }}a\neq b,\
b\in fn(E)-\widetilde{c}.$

Since $M\stackrel{(\nu \widetilde{c})\overline{a}\langle E\rangle }{%
\longrightarrow }M^{\prime }$ and $a\neq b,\ b\in fn(E)-\widetilde{c},$ we
have $T^{PS}(M)\rightarrow \langle \overline{a}\langle \widetilde{c}%
\circledR T^{PS}(E)\rangle \rangle T^{PS}(M^{\prime })$ and $a\neq b\wedge
\wedge _{i=1}^nb\neq c_i\wedge \ (B\rightarrow \neg (\ominus b)\top ).$ By
the axiom $(a\neq b\wedge \wedge _{i=1}^nb\neq c_i\wedge \ (E\rightarrow
\neg (\ominus b)\top )\wedge ((\tilde \ominus )E\leftrightarrow
E))\rightarrow (b\circledR \langle \overline{a}\langle c_1\circledR ...c_n%
\circledR T^{PS}(E)\rangle \rangle \linebreak T^{PS}(M^{\prime })\rightarrow
\langle \overline{a}\langle b\circledR c_1\circledR ...c_n\circledR %
T^{PS}(E)\rangle \rangle T^{PS}(M^{\prime })),$ we have $T^{PS}(P)=b%
\circledR T^{PS}(M)\linebreak \vdash _{SL}(b\circledR \langle \overline{a}%
\langle c_1\circledR ...c_n\circledR T^{PS}(E)\rangle \rangle
T^{PS}(M^{\prime })\vdash _{SL}\langle \overline{a}\langle b\circledR c_1%
\circledR ...c_n\circledR T^{PS}(E)\rangle \rangle T^{PS}(M^{\prime }).$%
\TeXButton{End Proof}{\endproof}

\begin{center}
{\bf Appendix E. Proof of Proposition 11}
\end{center}

{\bf Proposition} {\bf 11 }For any $A\in WL,$ $T^{WM}(A)\in \mu SL;$ for any
$P\in Pr,$ $P\models _{\mu SL}T^{WM}(A)\Leftrightarrow P\models _{WL}A.$

$Proof:$ We only discuss the case $A=\langle \langle a\langle A_1\rangle
\rangle \rangle A_2,$ other cases are similar.

Suppose $P\models _{\mu SL}T^{WM}(A).$ Since $[[\mu X.C(X)]]_{Pr}^e=\cup
_i[[C^i(\bot )]]_{Pr}^e,$ if $P\linebreak \in [[\mu X.C(X)]]_{Pr}^e,$ then $%
P\in [[C^i(\bot )]]_{Pr}^e$ for some $i$. Let $B=\langle a\langle
T^{WM}(A_1)\rangle \rangle \linebreak (\mu Y.(T^{WM}(A_2)\vee \langle \tau
\rangle Y)),$ then $P\models _{\mu SL}$ $B\vee \langle \tau \rangle B\vee
\langle \tau \rangle \langle \tau \rangle B...\vee \langle \tau \rangle ^iB,$
here $\langle \tau \rangle ^{i+1}B$ denotes $\langle \tau \rangle (\langle
\tau \rangle ^iB),$ $\langle \tau \rangle ^0B$ is $B.$ Hence $P\stackrel{%
\varepsilon }{\Longrightarrow }Q,$ $Q\in [[\langle a\langle
T^{WM}(A_1)\rangle \rangle \linebreak (\mu Y.(T^{WM}(A_2)\vee \langle \tau
\rangle Y))]]_{Pr}^e.$ Hence $Q\stackrel{a\langle E\rangle }{\longrightarrow
}Q^{\prime },$ $E\in [[T^{WM}(A_1)]]_{Pr}^e,$ and $Q^{\prime }\in [[\mu
Y.(T^{WM}(A_2)\vee \langle \tau \rangle Y)]]_{Pr}^e.$ By the similar
discuss, we have that $Q^{\prime }\stackrel{\varepsilon }{\Longrightarrow }%
Q^{\prime \prime }$ and $Q^{\prime \prime }\in [[T^{WM}(A_2)]]_{Pr}^e.$
Hence $P\stackrel{a\langle E\rangle }{\Longrightarrow }Q^{\prime \prime },$ $%
E\in [[T^{WM}(A_1)]]_{Pr}^e,$ and $Q^{\prime \prime }\in
[[T^{WM}(A_2)]]_{Pr}^e.$ We have $P\models _{WL}A.$ The converse claim is
similar. \TeXButton{End Proof}{\endproof}

\begin{center}
{\bf Appendix F. Proof of Proposition 12}
\end{center}

{\bf Proposition} {\bf 12 }$\vdash _{\mu SL}A|!A\leftrightarrow !A$

$Proof:$ Since by the inference system, $\vdash _{\mu SL}S(\mu
X.S(X))\rightarrow \mu X.S(X),$ we have $\neg \mu X.S(X)\rightarrow \neg
S(\mu X.S(X)).$ Let $S(X)=\neg (A|\neg X),$ then $\neg \mu X.S(X)=\neg \mu
X.\neg (A|\neg X)=!A,$ $\neg S(\mu X.S(X))=A|\neg \mu X.\neg (A|\neg
X)=A|!A. $ Therefore we get $\vdash _{\mu SL}!A\rightarrow A|!A.$

Since by the inference system, $\vdash _{\mu SL}!A\rightarrow A|!A,$ we have
$\vdash _{\mu SL}\neg (A|A|!A)\rightarrow \neg (A|!A).$ Let $T(X)=\neg
(A|\neg X),$ then $T(\neg (A|!A))=\neg (A|A|!A).$ Since $\vdash _{\mu
SL}T(\neg (A|!A))\rightarrow \neg (A|!A),$ by the inference system, we have $%
\vdash _{\mu SL}\mu X.T(X)\rightarrow \neg (A|!A).$ Furthermore, $\mu
X.T(X)=\mu X.\neg (A|\neg X)=\neg !A,$ hence $\vdash _{\mu SL}\neg
!A\rightarrow \neg (A|!A),$ we have $\vdash _{\mu SL}A|!A\rightarrow !A.$
\TeXButton{End Proof}{\endproof}

\end{document}